\documentclass[aps,prl,twocolumn,showpacs,floatfix,superscriptaddress]{revtex4}
\usepackage{amsthm,amsmath,graphicx}

\begin{document}

\title{Fracture strength of disordered media: Universality, interactions and tail asymptotics}

\author{Claudio Manzato}
\affiliation{Dipartimento di Fisica, Universit{\`a} di Modena e
Reggio Emilia, 41100 Modena, Italy}
\author{Ashivni Shekhawat}
\affiliation{LASSP, Physics Department, Clark Hall, Cornell University, Ithaca, NY 14853-2501}
\author{Phani K.~V.~V.~Nukala}
\affiliation{Computer Science and Mathematics Division,
Oak Ridge National Laboratory, Oak Ridge, TN 37831-6164}
\author{Mikko J.~Alava}
\affiliation{Department of Applied Physics, Aalto University, School
of Science, PO Box 14100, FI-00076 Aalto, Finland}
\author{James P.~Sethna}
\affiliation{LASSP, Physics Department, Clark Hall, Cornell University, Ithaca, NY 14853-2501}
\author{Stefano Zapperi}
\affiliation{CNR - Consiglio Nazionale delle Ricerche, IENI, Via R. Cozzi 53, 20125 Milano, Italy  }
\affiliation{ISI Foundation, Viale S. Severo 65, 10133 Torino, Italy }

\begin{abstract}
We study the asymptotic properties of fracture strength distributions of 
disordered elastic media by a combination of renormalization group, 
extreme value theory, and numerical simulation. We investigate the 
validity of the `weakest-link hypothesis' in the presence of realistic 
long-ranged interactions in the random fuse model.
Numerical simulations indicate that the
fracture strength is well described by the Duxbury-Leath-Beale (DLB)
distribution which is shown to flow asymptotically to the Gumbel distribution. 
We explore the relation between the extreme value distributions and the DLB type asymptotic distributions, 
and show that the universal extreme value forms may not be 
appropriate to describe the non-universal low-strength tail.
\end{abstract}

\pacs{62.20.mj,62.20.mm,62.20.mt,64.60.ae,02.50.-r,64.60.Q-}


\maketitle

\date{\today}
It has been known for centuries that larger bodies have lower
fracture strength. The traditional explanation of this size effect
is the `weakest link' hypothesis:  the sample is envisaged as a set
of non-interacting sub-volumes with different failure thresholds, and
its strength is determined by the failure of the weakest region. If
the sub-volume threshold distribution has a power law tail near zero then the
strength distribution can be shown to converge to the universal
Weibull distribution for large sample sizes \cite{weibull1939}, an
early application of extreme value theory (EVT) \cite{gumbel1958}.

Often failure occurs due to the presence and growth of micro-cracks
whose long-range interactions call the notion of independent
sub-volumes into question. There have been two broad approaches to
address such interactions: fiber bundle models and fracture network
models \cite{mikko2006}. Fiber bundles transfer load by various rules
as individual fibers fail; in some particular cases exact asymptotic
results for the failure distribution have been derived
\cite{phoenix2000}, 
and do not explicitly fall into any
of the extreme value statistics universal forms. 
Fracture network models
consider networks of elastic elements with realistic long-range
interactions and disorder. A particularly simple approach is based
on the random fuse model (RFM) \cite{arcangelis85,mikko2006}, where
one approximates continuum elasticity with a discretized scalar
representation.
It has been suggested that in the weak disorder limit, fracture would be
ruled by the longest micro-crack present in the system
\cite{freudenthal1968,duxbury1987,beale1988,chakrabarti1997}. 
By using critical droplet theory type arguments, one can 
show that an 
exponential distribution of micro-cracks leads to the
DLB distribution of failure strengths
\cite{duxbury1987}, which again does not explicitly have an extreme
value form.

These studies raise three important questions. 
First, what is
the importance of elastic interactions in determining the strength
distributions, and does the weakest link hypothesis hold in presence of such interactions? 
Second, what is the
relation between the DLB type asymptotic strength distributions and the
universal forms predicted by EVT? 
Third, how should one best extrapolate from measured
strength distributions to predict the probability of rare catastrophic
events? We use renormalization group (RG) ideas, EVT, and
simulations of the two dimensional RFM to explore these questions.
We conclude that (i) the weakest link hypothesis is valid for
large samples even in the presence of long-ranged elastic interactions, (ii) the
asymptotic forms of the strength distribution for these interacting
models is compatible in disguise with EVT, but of
the Gumbel form rather than the Weibull form, and, (iii) the use of extreme value
distributions to estimate the probability of rare
events, though common in the experimental literature, is not always justified
theoretically. DLB type asymptotic distributions (or those derived by Phoenix \cite{phoenix2000})
which depend on the details of the material are
necessary to safely extrapolate deep into the tails of the failure
distribution.

The RG and the EVT present two equivalent, yet contrasting,
approaches to the study of the universal aspects of extreme value distributions in general~\cite{gyorgyi2010}, and fracture strengths in particular.
The natural framework to investigate the role of interactions and the corrections 
to scaling that emerge as the system size is changed is provided by the RG theory.  
In contrast, the EVT facilitates the study of domains of attraction and convergence issues.
The non-universal, yet important, behavior of the low reliability tail of the distribution
is not described adequately by either the RG or the EVT. To study such non-universal 
features one needs to develop DLB type asymptotic theories.  

Typically, a
RG transformation proceeds in two steps: in the first step the system is
coarse-grained by eliminating short length-scale degrees of freedom,
and then the resulting system is rescaled. The RG coarse-graining 
for fracture is equivalent to the weakest link hypothesis: a system of size $L$ in $d=2$
dimensions survives at a stress $\sigma$ if its 4 ($=2^d)$ sub-systems of size $L/2$ survive at the same stress. This coarse-graining leads to the 
following recursion relation for $S_{L}(\sigma)$ --- the probability that a system of size $L$ does not fail under a
stress $\sigma$:
\begin{equation}
S_{L}(\sigma)=\left[S_{L/2}(\sigma)\right]^4.
\label{eq:RG}
\end{equation}
The second step of the RG transformation is to rescale the stress
suitably and look for a fixed point distribution
$S^*$ that is invariant under RG
\begin{equation}
S^*(\sigma)={\cal R}[S^*(\sigma)]=\left[S^*(a\sigma + b)\right]^4.
\label{eq:RGFixedPT}
\end{equation}
Instead of applying Eq.~\ref{eq:RG} iteratively like the RG, the EVT formulation consider the large length-scale limit directly
\begin{equation}
S^*(\sigma) =  \lim_{L\to\infty}[S_{L_0}(A_L\sigma + B_L)]^{(L/L_0)^d},
\label{eq:EVT}
\end{equation}
where $L_0$ is a characteristic length-scale.
The functional equations~\ref{eq:RGFixedPT}, \ref{eq:EVT} are known to have only three solutions: the Gumbel, 
the Weibull, and the Fr\'{e}chet distributions. Of these, only the Gumbel 
($S^*(\sigma) = \Lambda(\sigma) \equiv \exp[-e^{\sigma}],\ \sigma \in \Re,\ a = 1,\ b = \log 4$) and the Weibull 
($S^*(\sigma) = \Psi_\alpha(\sigma)\equiv e^{-\sigma^\alpha},\ \sigma,\alpha > 0,\ a = 4^{(-1/\alpha)},\ b = 0$) distributions
are relevant for fracture
. The large length norming constants, $A_L,\ B_L$, satisfy the following
asymptotic relations $A_{2L}/A_L \to 1/a,\ |B_{2L} - B_L|/A_L \to b/a$.
\par
To test the validity of the weakest link hypothesis (Eq.~\ref{eq:RG}) in presence of  long-range
elastic interactions, we perform large scale simulations
of the RFM \cite{arcangelis85,mikko2006}, considering a tilted
square lattice (diamond lattice) with $L\times L$ bonds of unit
conductance. Initially we remove a fraction $1-p$ of the fuses at
random, where $p$ is varied between $1-p=0.05$ and $1-p=0.35$ (the
percolation threshold for this model is at $p=1/2$). Periodic
boundary conditions are imposed in the horizontal direction and a
constant voltage difference, $V$, is applied between the top and the
bottom of lattice system bus bars. The Kirchhoff equations are
solved to determine the current distribution on the lattice. A fuse 
breaks irreversibly whenever the local current exceed a threshold
that we set to one. Each time a fuse is broken, we re-calculate the
currents in the lattice and find the next fuse to break. 
The process is repeated until the system is disconnected. In
the present simulations, we have considered system sizes from $L=16$
to $L=1024$ and various values of $p$. 
To explore the low strength tail which is beyond the accessible range of most experiments, we typically average our
results over $10^5$ realizations of the initial disorder. The fuse model is equivalent to a scalar elastic
problem. Using this equivalence, the strain is defined as
$\epsilon=V/L$ and the stress is given by $\sigma=I/L$, where $I$ is
the current flowing in the lattice. The fracture strength is defined
as the maximum value of $\sigma$ during the simulation.

\begin{figure}[tpb]
\begin{center}
  \includegraphics[width=0.9\linewidth]{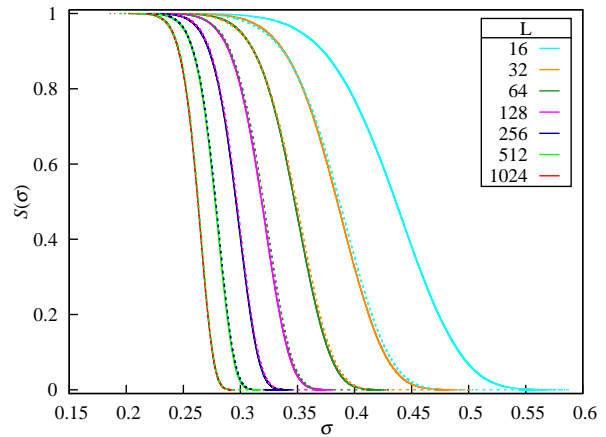}
\end{center}
\caption{Testing the weakest link hypothesis. Comparing the survival
probability $S_{L}(\sigma)$ for a $L\times L$ network (solid lines)
with that predicted by the weakest link hypothesis, $S_{L/2}(\sigma)^4$, 
(dotted lines) for $1-p = 0.10$. Note the excellent agreement
even for moderate system sizes.} \label{fig:1}
\end{figure}
The RG coarse-graining step (Eq.~\ref{eq:RG}) produces a natural 
test for the weakest link hypothesis.
In Fig.~\ref{fig:1} we report the survival probability 
$S_{L}(\sigma)$ for different system sizes $L$, compared
with those for systems of size $L/2$, rescaled according
to Eq.~\ref{eq:RG}. The agreement between the two
distributions is almost perfect for $L/2\geq32$, indicating that 
Eq.~\ref{eq:RG} is satisfied asymptotically. Corrections to scaling due
to the effect of distant micro-cracks are expected to decay as
$1/L^2$, as can be shown by a direct calculation, but are too small
for us to detect in simulations (Fig.~\ref{fig:1}).
We also tested wide rectangular systems with $L_x =2L_y$, finding
larger corrections, scaling roughly as $1/L$, which are still
irrelevant in the large system size limit.

\begin{figure}[tpb]
\begin{center}
\includegraphics[width=8cm]{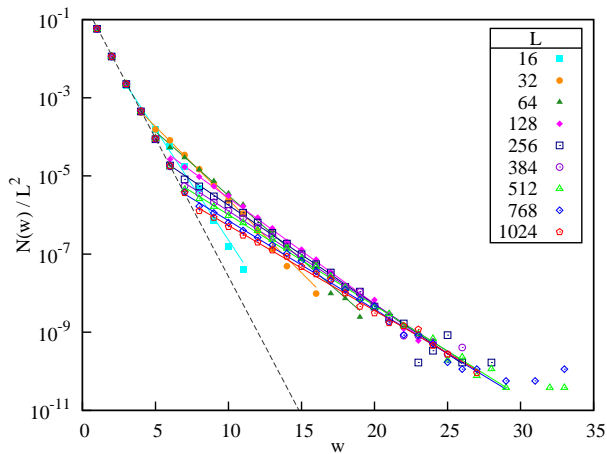}
\end{center}
\caption{Crack width distributions at peak load, $1-p = 0.10$. The initial
distribution of micro-crack widths ($N(w)$ is the number of clusters
of width $w$), is exponential with (dotted line, slope  $\approx -\log 2(1-p)$). As the system is
loaded, a few bonds break before catastrophic failure; these bonds
usually connect smaller clusters, producing extra cracks at large
widths. The resulting crack width distribution at the peak load
exhibits a size-dependent crossover to a different exponential
slope. Solid lines represent fits to an exponential.} \label{fig:2}
\end{figure}

\begin{figure}[tpb]
\begin{center}
(a)\includegraphics[width=7.5cm]{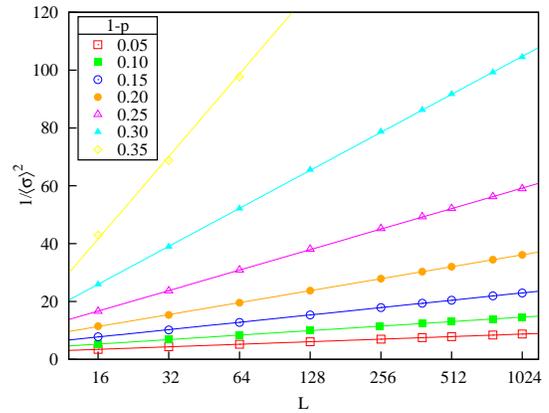}
\end{center}
\begin{center}
(b)\includegraphics[width=7.5cm]{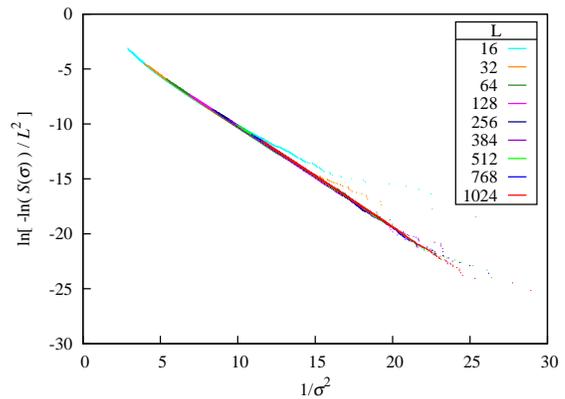}
\end{center}
\caption{Testing the DLB distribution of failure stresses.
(a)  The average failure stress as a function of system size $L$ at
various bond fractions $p$ (symbols) can be fit well by the
DLB form (solid lines), except close to the percolation
threshold ($1-p > 0.3$).  (b) A collapse of the strength
distribution for different system sizes at $1-p=0.1$, such that the
DLB form would collapse onto a straight line.}
\label{fig:3}
\end{figure}

Duxbury {\em et al.}~related the survival distribution to the
distribution of micro-crack widths $w$ \cite{duxbury1987}. At the
beginning of the simulation the `per-site' probability distribution of a crack of width $w$ is  $P(w <
w') = 1 - e^{-w'/w_0}$, where $w_0 \sim -1/\log 2(1-p)$ \cite{footnote2}.
Hence, the distribution of the longest crack, $w_m$, in a lattice with $L^2$ sites is given by
\begin{equation}
    P(w_m < w')  = \left(1 - e^{-w'/w_0}\right)^{L^2}.
\end{equation}
The stress at the tip of a crack
of width $w$ is asymptotic to $\sigma K \sqrt{w}$, where $\sigma$ is the
applied far-field stress, and $K$ is a lattice dependent constant. 
A sample survives until the largest crack becomes unstable
when its tip stress reaches a threshold $\sigma_{th} = \sigma K
\sqrt{w}$. Therefore, we have
\begin{equation}
 S_L(\sigma)\simeq\left(1 - e^{- (\frac{\sigma_0}{\sigma})^2}\right)^{L^2}
 \simeq D_L(\sigma),
\label{eq:DLB}
\end{equation}
where $\sigma_0 \equiv \sigma_{th}/K\sqrt{w_0}$ and $D_L(\sigma)\equiv \exp[-L^2
e^{-(\sigma_0/\sigma)^2}]$ is the DLB distribution. To apply the above derivation to the
failure stress, we  first check the distribution of micro-crack
lengths at peak load.
As shown in Fig.~\ref{fig:2}, the
distribution is exponential, but due to damage accumulation, the
slope of the tail changes with respect to the initial distribution.
This appears to be due to bridging events in which two neighboring
cracks join, leading to a modification of Eq.~\ref{eq:DLB} as
discussed in Ref.~\cite{duxbury1987}. 
Thus, damage accumulation, though very small, is
relevant because it changes the exponent of the micro-crack distribution. 
The exponential form  of the crack length
distribution tail, however, suggests that the DLB form should still be
valid, as demonstrated in Fig.~\ref{fig:3}. In particular, the
average failure stress scales as $\langle \sigma \rangle =
\sigma_0/\sqrt{\log (L^2)}$ (Fig.~\ref{fig:3}a) and the
distributions for different $L$ all collapse into a straight line
when plotted in terms of rescaled coordinates (Fig.~\ref{fig:3}b).

Our arguments thus far are seemingly paradoxical. On the one hand we have argued on very general grounds that 
the distribution of failure strengths must be either Gumbel or Weibull, while on the other hand we have 
checked that the failure distribution for fuse-networks is of the rather different form proposed by Duxbury \emph{et~al.} How can 
this `paradox' be resolved? It is easy to check that the DLB distribution, when 
rescaled and centered properly, yields a Gumbel distribution, i.e., 

\begin{equation}
\lim_{L\to\infty} D_L(A_L\sigma + B_L) = \Lambda(\sigma),
\label{eq:DuxburyToGumbel}
\end{equation}
as can be demonstrated by a straightforward calculation using $A_L =
\sigma_0/(2 (\log(L^2))^{3/2})$  and $B_L =
\sigma_0/\sqrt{\log( L^2)}$. 
The above result is striking because fracture distributions are 
usually assumed to not be of the Gumbel form, since fracture must happen at positive stress, 
while the Gumbel distribution has support for negative arguments as well.
This is akin to arguing that the normal distribution is not valid for test scores since
scores must always be positive. Nonetheless, it brings us to the issue of convergence and validity of extreme value distributions as opposed to 
DLB type distributions.
\par
The extreme value distributions, $S^*(\sigma)$ (=$\Lambda(\sigma)$ or $\Psi_\alpha(\sigma)$), 
are a uniform approximation to the 
true survival function, $S_L(\sigma)$, for all $\sigma$ in the limit of large $L$, i.e.,
\begin{equation}
\lim_{L\to\infty}\left(\sup_{\sigma \in \Re} \left| S_L(\sigma) - S^*\left(\frac{\sigma - B_L}{A_L}\right)\right|\right)  = 0.
\label{eq:UniformValidity}
\end{equation}
In contrast, DLB type distributions~\cite{footnote3}, are based on material details, and are asymptotically correct in the low reliability tail, i.e.,
\begin{equation}
\lim_{L\to\infty}\left(\lim_{\sigma\to 0} \frac{ 1-D_L(\sigma)}{1-S_L(\sigma)}\right) = 1.
\label{eq:ExactTail}
\end{equation}

Note that the uniform convergence in Eq.~\ref{eq:UniformValidity} does not bound the \emph{relative} error in the low 
reliability tail, while the asymptotic convergence in Eq.~\ref{eq:ExactTail} does.

The above discussion hints at an underlying question: How to accurately predict the 
probability of rare small-strength events with limited experimental data? 
The standard practice is to measure the failure distribution of construction
beams or micro-circuit wires, fit to the universal Weibull or Gumbel form, and
extrapolate. However, as we have argued, this approach
can be dangerous.
The low reliability tail is non-universal, and must be modeled by a theory that, like DLB, 
 accounts for microscopic details.
Such theories, analogous to critical droplet theory (low temperatures), instantons
(low $\hbar$), and Lifshitz tails (low disorder, deep in the band
gap) are by construction accurate in the low reliability tail.
It is interesting to observe that usually the RG and the critical droplet theory
address continuous and abrupt phase transitions, respectively, yet here
these two approaches both apply to fracture.
\par
The convergence to extreme value distributions can 
be extremely slow. For the RFM, let $z$ be number of standard deviations 
up to which the Gumbel approximation is accurate within a relative error of $\epsilon$.
By using the Edgeworth type expansions 
for the extreme value distributions~\cite{haan1996}, we find 
\begin{displaymath}
\frac{z\pi}{\sqrt{6}} = \left\{ \begin{array}{ll}
		\sqrt{\eta}\exp[-\frac{\sqrt{\eta}}{2}\exp[-\frac{\sqrt{\eta}}{2}\exp[ \ldots ]     ]   ],		& \eta < 4 e^2 \\
		\log \eta - 2 \log[\log\eta -2\log[\ldots       ]         ],		& \eta > 4 e^2,
		\end{array}\right.
\end{displaymath}
where the ellipsis indicate an infinite recursion, and $\eta = -(4/3)\log(1-\epsilon)\log(L^2)$. For an accuracy of 10\% 
at one standard deviation a sample volume of $L^2\approx 10^{18}$ is required, while at 2 standard deviations the 
required sample volume is about $L^2\approx 10^{264}$. As a comparison, for the Gaussian approximation to the 
mean of a sample of $M (\gg 1)$ random variables (normalized so that $E[X] = 0,\ E[X^2] = 1,\ E[X^3] = \gamma$) we get, 
$z \sim \Delta^{1/3} + \Delta^{-1/3} + \mathcal{O}(\Delta^{-4/3})$, where $\Delta = 6\epsilon \sqrt{M}/\gamma$, 
thus $z \approx 3$ for $\epsilon = 0.1,\ M = 3000,\ \gamma = 2$, where the value $\gamma = 2$ corresponds to the standard 
exponential distribution. However, the 
universal extreme value forms are not always dangerous for extrapolation. One can show that they are valid asymptotic forms, 
\`{a} la Eq.~\ref{eq:ExactTail}, if they satisfy the condition of tail equivalence~\cite[p.~102]{resnick1987extreme}\cite{anderson1978}:
\begin{equation}
\lim_{\sigma\to 0} \frac{ 1 - S_L(\sigma)}{1 - S^*(\sigma)} = C,\ 0 < C < \infty.
\label{eq:tailEquivalence}
\end{equation}
The success of the classical example of a Weibull distribution of failure strengths emerging from a power-law micro-crack length distribution
may be due to the tail equivalence of the microscopic and the Weibull distributions.

In conclusion, by using a combination of renormalization group, extreme value theory, and numerical simulations we
have shown that the failure strength of an elastic solid with a
random distribution of micro-cracks follows the DLB distribution which 
asymptotically falls into the Gumbel universality class. 
The non-universal low reliability tail of the strength distribution
may not be described by the universal extreme value distributions, and thus the 
common practice of fitting experimental data to universal forms and extrapolating in the tails 
is questionable. Theories that account for 
microscopic mechanisms of failures, the DLB distribution for instance, 
are required for accurate prediction of low strength failures.
In our study the emergence of a Gumbel distribution of fracture strengths 
is surprising, and brings into question the widespread use of 
the Weibull distribution for fitting experimental data.

We thank Sidney I.~Resnick, Leigh Phoenix, and Bryan Daniels for insightful discussions. 
We acknowledge support from DOE-BES DE-FG02-07ER-46393 (AS and JPS), the Academy of 
Finland via the Center of Excellence program (MJA), the ComplexityNet pilot project LOCAT (SZ),  the HPC-EUROPA2 project (228398)
supported by the European Commission - Capacities Area - Research Infrastructures, and the DEISA Consortium
(EU projects FP6 RI-031513 and FP7 RI-222919) within the DEISA Extreme Computing Initiative.



\begin{thebibliography}{15}
\bibitem[{foo({\natexlab{a}})}]{footnote2}
\bibinfo{note}{The factor 2 in $w_0 \sim -1/\log2(1-p)$ is due to the fact
  that, on the diamond lattice, there are $2^w$ crack `backbones' of width $w$
  with a given site as their left end.}
\bibitem[{foo({\natexlab{b}})}]{footnote3}
\bibinfo{note}{The DLB distribution is asymptotically exact in the limit of
  $1-p \to 0$.}
\end{thebibliography}

\end{document}